\documentclass[aps,prl,onecolumn,preprintnumbers,superscriptaddress,nofootinbib,floatfix,10pt]{revtex4}

\usepackage{amsfonts,amssymb,stmaryrd,latexsym,amsmath,braket}
\usepackage{graphicx,subfigure,blindtext}
\usepackage{bm}
\usepackage{bookmark}
\usepackage{times}
\usepackage{slashed}

\begin{document}

\title{Essential elements for nuclear binding}

\author{Bing-Nan~Lu}
\affiliation{Facility for Rare Isotope Beams and Department of Physics and
Astronomy,
Michigan State University, MI 48824, USA}

\author{Ning~Li}
\affiliation{Facility for Rare Isotope Beams and Department of Physics and
Astronomy,
Michigan State University, MI 48824, USA}

\author{Serdar~Elhatisari}
\affiliation{Helmholtz-Institut f{\"u}r Strahlen- und Kernphysik and Bethe
Center
for Theoretical Physics, Universit\"at Bonn, D
-53115 Bonn, Germany}
\affiliation{Faculty of Engineering, Karamanoglu Mehmetbey University,
Karaman 70100, Turkey}

\author{Dean~Lee}
\affiliation{Facility for Rare Isotope Beams and Department of Physics and
Astronomy,
Michigan State University, MI 48824, USA}

\author{Evgeny~Epelbaum}
\affiliation{Ruhr University Bochum, Faculty of Physics and Astronomy, Institute
for Theoretical Physics II,
D-44870 Bochum, Germany}

\author{Ulf-G.~Mei{\ss}ner}
\affiliation{Helmholtz-Institut f{\"u}r Strahlen- und Kernphysik and Bethe
Center
for Theoretical Physics, Universit\"at Bonn, D
-53115 Bonn, Germany}
\affiliation{Institute for Advanced Simulation, Institut f\"ur Kernphysik,
and
J\"ulich Center for Hadron Physics, Forschungszentrum J\"ulich,
D-52425 J\"ulich, Germany}
\affiliation{Tbilisi State  University,  0186 Tbilisi, Georgia}


\begin{abstract}
How does nuclear binding emerge from first principles?
Our current best understanding of nuclear forces is based on a systematic
low-energy expansion called chiral effective field theory.  However, recent
{\it ab initio} calculations of nuclear structure have  found that not all
chiral effective field theory interactions give accurate predictions with
increasing nuclear density.  In this letter we address the reason for this
problem and the first steps toward a solution. Using nuclear lattice simulations,
we deduce the minimal nuclear interaction that can reproduce the ground state
properties of light nuclei, medium-mass nuclei, and neutron matter simultaneously
with no more than a few percent error in the energies and charge radii. 
We find that only four parameters are needed.  With these four parameters
one can accurately  describe neutron matter up to saturation density and
the ground state properties of nuclei up to calcium.  Given the absence of
sign oscillations
in these lattice Monte Carlo simulations and the mild scaling of computational
effort scaling with nucleon number, this work provides
a pathway to high-quality simulations in the future with as many as one or
two hundred nucleons.  
\end{abstract}

\maketitle


Chiral effective field theory ($\chi$EFT) is a first principles approach
to nuclear forces where interactions are arranged as a low-energy expansion
in powers of momentum and pion mass~\cite{Epelbaum:2008ga,Machleidt:2011zz}.
 While
many calculations establish the reliability of $\chi$EFT in describing the properties of light
nuclei \cite{Hupin:2014iqa,Navratil:2007aj,Lonardoni:2017hgs,Piarulli:2017dwd,Epelbaum:2018ogq}, the binding energies and charge radii of medium mass nuclei are not consistently reproduced ~\cite{Wloch:2005za,Hagen:2008iw,Hagen:2012sh,Binder:2013xaa,Epelbaum:2013paa,Cipollone:2014hfa,Ekstrom:2015rta,Lonardoni:2017hgs}.  One well-known example is that the charge radius of $^{16}$O tends to be too small for most of interactions in the literature \cite{Wloch:2005za,Hagen:2012sh,Binder:2013xaa,Epelbaum:2013paa,Cipollone:2014hfa}.
The core issue is that $\chi$EFT many-body calculations do not yet
give reliable and accurate predictions at higher nuclear densities.
We note that there have been efforts to improve the convergence of many-body 
calculations by rearranging the chiral effective field theory expansion at nonzero density~\cite{Meissner:2001gz,Lacour:2009ej}.  If one reaches high 
enough orders in the $\chi$EFT expansion, then the systematic errors will 
eventually decrease as more and more low-energy parameters are tuned to 
empirical data.  However, the predictive power of the {\it ab initio} approach will be 
diminished as more data will be needed to constrain the higher-body 
forces.  Furthermore, the computational effort will increase 
significantly to the point where a first principles treatment may not be 
practical.

One pragmatic approach is to further constrain the nuclear force using nuclear
structure data from medium mass nuclei or the saturation properties of nuclear
matter \cite{Ekstrom:2015rta}. 
This approach has been applied successfully in several
recent calculations~\cite{Hagen:2015yea,Hagen:2016uwj,Simonis:2017dny,Morris:2017vxi}.  A rather different
 line of investigation has looked at the microscopic origins of the problem.
 In Ref.~\cite{Elhatisari:2016owd} numerical evidence is shown that nuclear matter sits
near a quantum phase transition between a Bose gas of alpha particles and
nuclear liquid.  It is argued that local SU(4)-invariant forces play an increasingly
important role at higher nuclear densities. The term local refers to velocity-independent 
interactions, as opposed to nonlocal interactions which are velocity dependent. The SU(4) 
refers to Wigner's approximate
symmetry of the nuclear interactions where the four nucleonic degrees of
freedom (proton spin-up, proton spin-down, neutron spin-up, neutron spin-down)
can be rotated into each other~\cite{Wigner:1936dx}.

The importance of SU(4)-invariant interactions can be understood in terms
of coherent enhancement.   Spin-dependent forces tend to cancel when summing
over all possible nucleonic spin configurations.  For example, we have seen in lattice calculations of binding energies for closed shell systems that the contribution from the repulsive P-wave channels often cancels most of the contribution from the attractive P-wave channels.  We note also the intriguing analysis in Ref.~\cite{CalleCordon:2009ps} which demonstrates the connection between quantum chromodynamics with a large number of colors to Wigner's SU(4) symmetry for the S-wave interactions and Serber symmetry for the P-wave interactions.
Similarly most isospin-dependent
forces tend to cancel in symmetric nuclear matter due to the equal number of 
protons and neutrons,   
the one notable exception being the Coulomb interaction. The idea of SU(4)
universality at large S-wave scattering length has a rich history in nuclear
physics.  It is well known that the Tjon line relating $^{3}$H
and $^{4}$He binding energies is a manifestation of universality in
nuclear systems~\cite{Platter:2004zs,Klein:2018lqz}. It has also been shown that $^{3}$H
and $^{4}$He  are characterized by universal
physics associated with the Efimov effect~\cite{Konig:2016utl,Kievsky:2015dtk}. The coherent
enhancement of SU(4)-invariant forces in the nuclear many-body environment
 suggests a possible resurgence of
SU(4) symmetry in heavier nuclei as well.  This idea inspired
the exploratory work in Ref.~\cite{Elhatisari:2017eno} on
the structure of nuclei up through oxygen using an SU(4)-invariant interaction.
This built upon previous work
in  Ref.~\cite{Elhatisari:2016owd} which showed that local SU(4)-invariant interactions
have a particularly
strong effect on nuclear binding.  The special role of local forces
has also been studied by looking at the effective interactions between
two bound dimers in a one-dimensional model~\cite{Rokash:2016tqh}.

In this work we attempt to tie all of the loose threads together.  We start
by acknowledging that not every $\chi$EFT interaction will  give well controlled
and reliable results for heavier systems. Additional ingredients are
needed to make sure that the convergence of higher-order terms is under control.
 In order to see what the essential elements might be, we take a constructive
reductionist approach and deduce the minimal nuclear interaction that can
reproduce the ground state
properties of light nuclei, medium-mass nuclei, and neutron matter simultaneously
with no more than a few percent error in the energies and charge radii.



We start with a simple SU(4)-invariant leading order effective field theory
without explicit pions (pion-less EFT) on a periodic $L^3$ cube with  lattice
coordinates $\bm{n}=(n_{x,}n_{y},n_{z})$. The Hamiltonian is
\begin{equation}
H_{{\rm SU(4)}}=H_{\rm free}+\frac{1}{2!}C_{2}\sum_{\bm{n}}\tilde{\rho}(\bm{n})^{2}+\frac{1}{3!}C_{3}\sum_{\bm{n}}\tilde{\rho}(\bm{n})^{3},\label{eq:HSU4}
\end{equation}
where $H_{\rm free}$ is the free nucleon Hamiltonian with nucleon mass
$m=938.9$~MeV.
 The density operator $\tilde{\rho}(\bm{n})$ is defined in the same manner
as
in Ref.~\cite{Elhatisari:2017eno},
\begin{equation}
\tilde{\rho}(\bm{n})=\sum_{i}\tilde{a}_{i}^{\dagger}(\bm{n})\tilde{a}_{i}(\bm{n})+s_{L}\sum_{|\bm{n}^{\prime}-\bm{n}|=1}\sum_{i}\tilde{a}_{i}^{\dagger}(\bm{n}^{\prime})\tilde{a}_{i}(\bm{n}^{\prime}),
\end{equation}
where $i$ is the joint spin-isospin index and the smeared annihilation and
creation operators are defined as
\begin{equation}
\tilde{a}_{i}(\bm{n})=a_{i}(\bm{n})+s_{NL}\sum_{|\bm{n}^{\prime}-\bm{n}|=1}a_{i}(\bm{n}^{\prime}).
\end{equation}
The summation over the spin and isospin implies that the interaction is SU(4)
invariant. The parameter $s_L$ controls the range of the local part of the
interaction, while $s_{NL}$ controls the range of the nonlocal part of
the interaction.
The parameters $C_{2}$ and $C_{3}$ give the strength of the two-body
and three-body interactions, respectively.

In this letter we use a lattice spacing $a=1.32$~fm, which corresponds
to a momentum cutoff $\Lambda=\pi/a\approx465$~MeV. The dynamics
with momentum $Q$ much smaller than $\Lambda$ can be well described
and residual lattice artifacts are suppressed by powers of
$Q/\Lambda$~\cite{Klein:2018iqa}.
In Ref.~\cite{Lu:2015riz} we showed that the nucleon-nucleon scattering phase
shift
can be precisely extracted on the lattice using the spherical wall
method. In this work we fix the two-body interaction by fitting the
scattering length $a_{0}$ and effective range $r_{0}$. In each instance we
calculate the $S$-wave phase shifts below relative
momentum $P_{{\rm rel}}\leq50$~MeV using the spherical wall method and calculate
fit errors by comparing results with the
effective range expansion.

For systems with more than three nucleons, we use auxiliary-field Monte Carlo
lattice simulations for a cubic periodic box with length $L$~\cite{Lee:2008fa,Lahde:2019npb}.
 For nuclei with $A<30$ nucleons, we take $L\geq8$,
with larger values of $L$ for cases where more accuracy is desired.
For nuclei with $A\geq30$ we take $L=9$. The temporal lattice spacing
is 0.001~MeV$^{-1}$ and the projection time is set to 0.3~MeV$^{-1}$.
We find that these settings are enough to provide accurate results for systems
with $A\leq48$. We also use the recently-developed pinhole algorithm~\cite{Elhatisari:2017eno}
in order to calculate density distributions and charge radii.


We use few-body data with $A \le 3$ to fix the interaction coefficients $C_{2}$
and $C_{3}$, while the range of the interactions are controlled by the parameters 
$s_{NL}$ and $s_{L}$. The particular combination of $s_{NL}$ and $s_{L}$ 
we choose is set through a procedure we now describe.  In the few-body sector, the two smearing
parameters  $s_{NL}$ and $s_{L}$ produce very similar effects and are difficult
to distinguish from few-body data alone  \cite{Elhatisari:2016owd}. Therefore the chosen
values for $s_{NL}$ and $s_L$ are fixed later after calculating heavier nuclei.
The two-body interaction strength $C_{2}$ and interaction range are determined
by fitting the scattering length $a_{0}$ and
effective range $r_{0}$ averaged over the
two $S$-wave channels, $^{1}S_{0}$ and $^{3}S_{1}$. We adjust  $a_{0}$ to
minimize the corrections to the $^{3}$H and $^{4}$He binding energies that
arise from the differences between the two $S$-wave channels.  This process
gives an optimal value of $a_{0}=9.1$~fm, and we use this value for $a_{0}$
in what follows. We note that our SU(4)-invariant deuteron is degenerate
with the di-neutron ground state and has less than half of the physical deuteron
binding energy.  However this issue is easily fixed when SU(4)-breaking interactions
are included.
 For the SU(4)-averaged effective range we use $r_{0}=(r_{0}(^{1}S_{0})+r_{0}(^{3}S_{1}))/2\approx2.2$~fm.

We determine the three-body
coupling strength $C_{3}$  by fitting to the
$^{3}$H binding energy. At the physical point $B(^{3}$H$)=8.48$~MeV, the
 $^{4}$He binding energy
with the Coulomb interaction included is $28.9$~MeV.  This
is close to the experimental value $B(^{4}$He$)=28.3$~MeV.
We carry out this fitting process for several different pairs of  values
for $s_{NL}$ and $s_L$, and for each pair we calculate a handful of nuclear
ground states using auxiliary-field lattice Monte Carlo simulations.  As
described in the Supporting Online Materials section, we find that the pair $s_{NL}=0.5$ and
$s_L=0.061$
gives the best overall description.
The full set of optimized parameters are $C_2 = -3.41\times 10^{-7}$ MeV$^{-2}$,
$C_3 = -1.4\times 10^{-14}$ MeV$^{-5}$, $s_{NL}=0.5$, and $s_L=0.061$.

\begin{table}[htbp]
\centering
\begin{tabular}{c|cccc|ccc}
\hline\hline
 & $B$ & Exp.  & Coulomb & $B$/Exp. & $R_{{\rm ch}}$ & Exp. & $R_{{\rm ch}}$/Exp.  \tabularnewline
\hline
$^{3}$H & 8.48(2)(0) & 8.48 & 0.0 & 1.00 & 1.90(1)(1) & 1.76 & 1.08 \\
$^{3}$He & 7.75(2)(0) & 7.72 & 0.73(1)(0) & 1.00 &  1.99(1)(1) & 1.97 & 1.01\\
$^{4}$He & 28.89(1)(1) & 28.3 & 0.80(1)(1) & 1.02 & 1.72(1)(3) & 1.68 & 1.02 \\
$^{16}$O & 121.9(1)(3) & 127.6 & 13.9(1)(2) & 0.96 & 2.74(1)(1) & 2.70 & 1.01\\
$^{20}$Ne & 161.6(1)(1) & 160.6 & 20.2(1)(1) & 1.01 & 2.95(1)(1) & 3.01 & 0.98\\
$^{24}$Mg & 193.5(02)(17) & 198.3 & 28.0(1)(2) & 0.98 & 3.13(1)(2) & 3.06 & 1.02\\
$^{28}$Si & 235.8(04)(17) & 236.5 & 37.1(2)(3) & 1.00 & 3.26(1)(1) & 3.12 & 1.04\\
$^{40}$Ca & 346.8(6)(5) & 342.1 & 71.7(4)(4) & 1.01 & 3.42(1)(3) & 3.48 & 0.98\\
\hline
\end{tabular}

\caption{{\bf Comparison of calculations and experiments for selected nuclei.}
The calculated binding energies
and charge radii of $^{3}$H, $^{3}$He and selected alpha-like nuclei compared
with experimental values. The Coulomb interaction is taken into
account perturbatively. The first and second parentheses denote the Monte
Carlo
error and time extrapolation error, respectively.
All energies are in MeV and all radii in fm. Experimental binding energies
are taken from Ref.~\cite{Wang:2017} and radii from Ref.~\cite{Angeli:2013epw}.}
\label{tab:The-calculated-bindings-radii}
\end{table}

In Table~\ref{tab:The-calculated-bindings-radii} we show the binding energies
and charge radii for selected nuclei. For comparison
we also list the experimental values and the calculated Coulomb energy.
While the $^{3}$H energy is exact due to the fitting procedure, all
the other values are predictions. The largest relative error in binding
energy is 4\% and occurs for $^{16}$O.
 The largest relative
error in the charge radius is 8\% and occurs for $^{3}$H.  For the
calculations of nuclear charge radii, we have taken into account the charge
radius of the proton.

\begin{figure}[!ht]
\centering
\begin{centering}
\includegraphics[width=16cm]{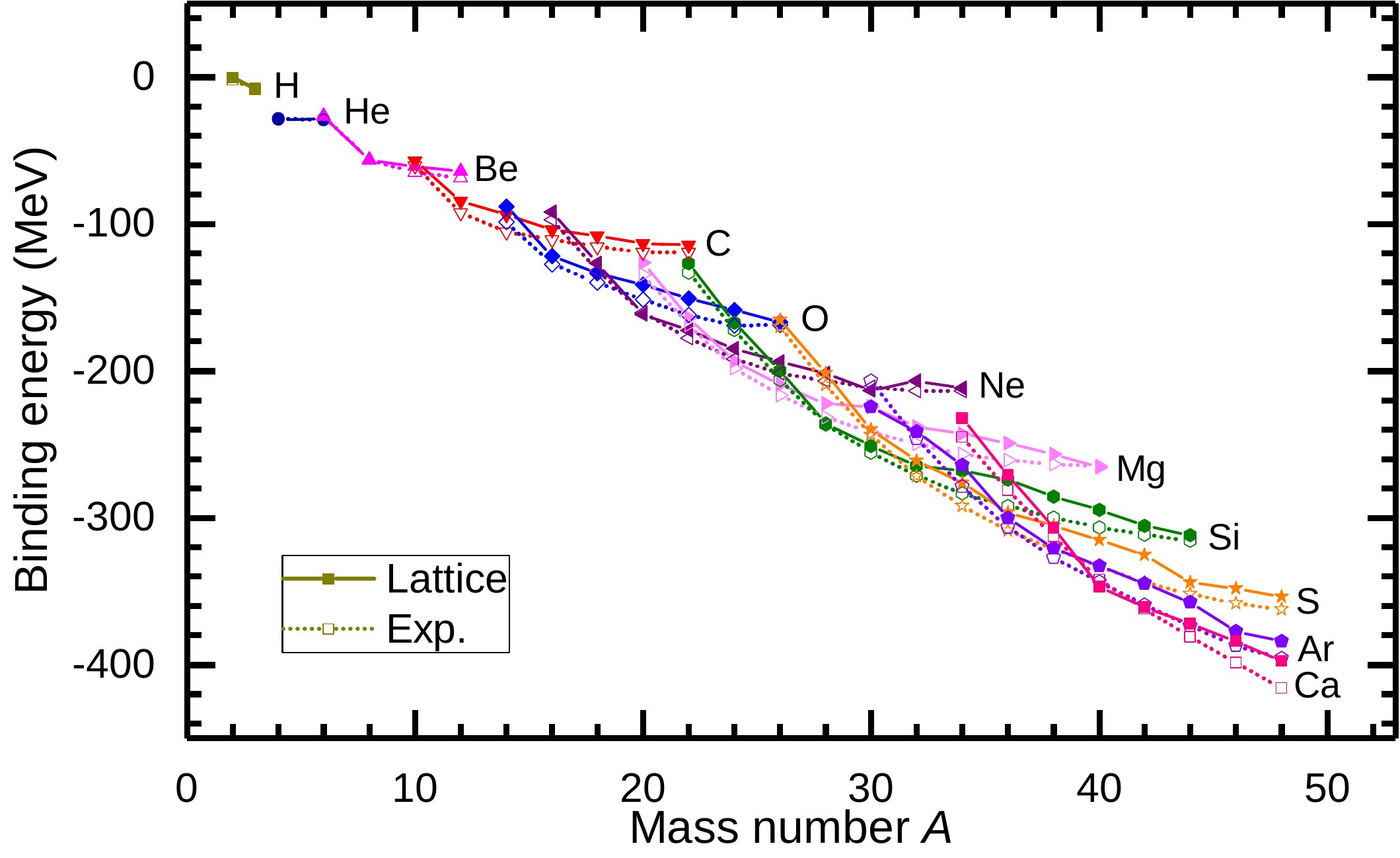}
\par\end{centering}
\caption{{\bf Nuclear binding energies.} The calculated binding energies
from $^{3}$H to $^{48}$Ca. The solid symbols denote the lattice
results and the open symbols denote the experimental values. Different
symbols and colors denote different element. The Coulomb interaction
is taken into account perturbatively. The experimental values are taken from
Ref.~\cite{Wang:2017}.}
\label{fig:The-calculated-bindings}
\end{figure}

We now calculate the binding
energies for 86 bound even-even nuclei (even number of protons, even number
of neutrons) with up to $A=48$ nucleons. The results are shown and compared
with empirical data in Fig.~\ref{fig:The-calculated-bindings}.
Because the interaction has an exact SU(4) symmetry, we are free of
the sign problem and can calculate the binding energies with high
precision. In Fig.~\ref{fig:The-calculated-bindings} all of the Monte
Carlo error bars are smaller than the size of the symbols. The remaining errors 
due to imaginary time and volume extrapolations are also small, less than 1\% relative 
error, and thus are also not explicitly shown. 
In Fig.~\ref{fig:The-calculated-bindings} we see that the general trends
for the binding energies along each isotopic chain are well reproduced.
In particular, the isotopic curves on the proton-rich
side are close to the experimental results. The discrepancy
is somewhat larger on the neutron-rich side and is a sign of missing effects
such as spin-dependent interactions.
%

\begin{figure}[!ht]
\centering
\begin{centering}
\includegraphics[width=12cm]{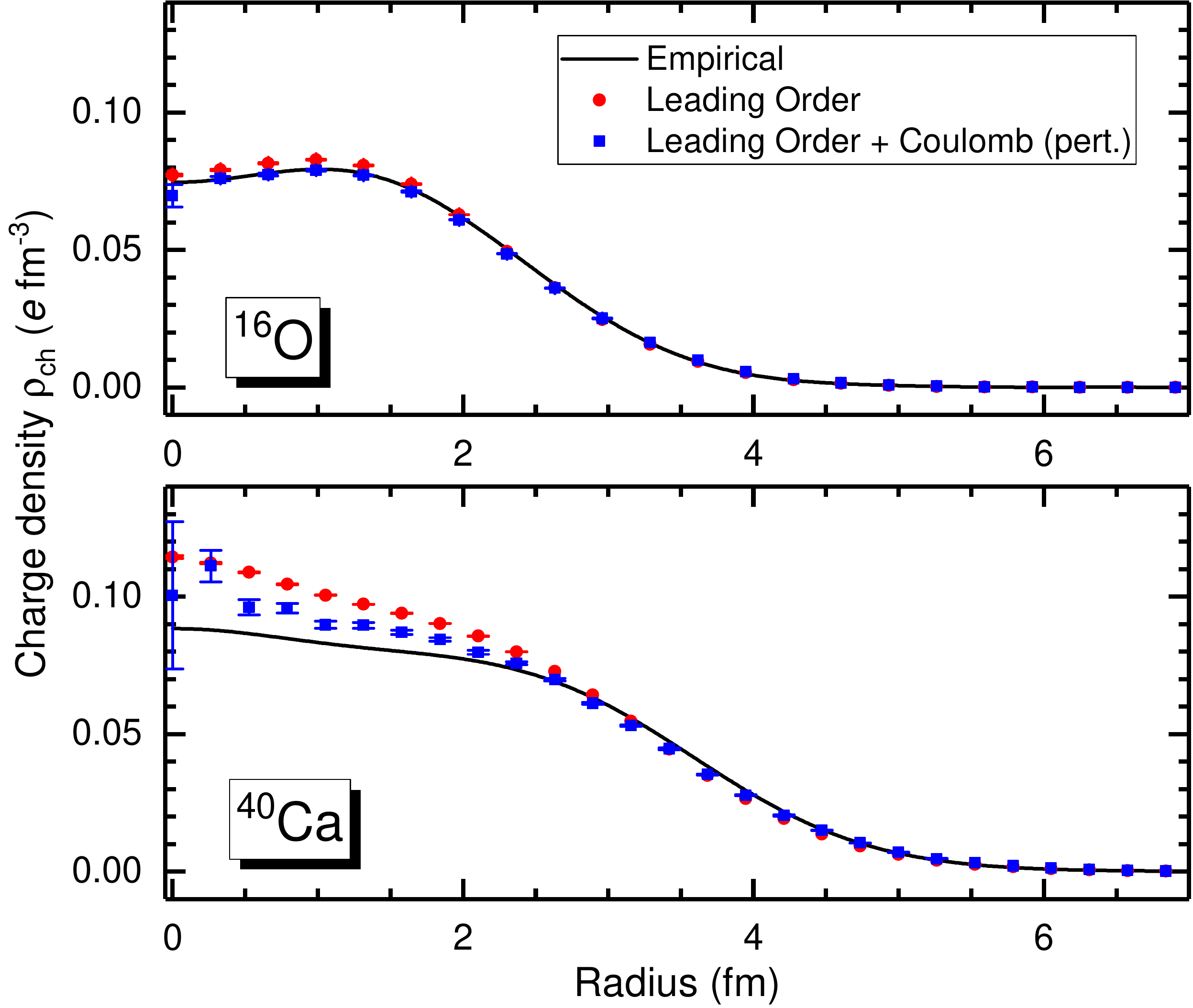}
\par\end{centering}
\caption{{\bf Charge density distributions.} The calculated $^{16}$O
  and $^{40}$Ca charge densities compared with the empirical results.
The circles denote the results without Coulomb interaction. The squares
denote the results with the Coulomb interaction included perturbatively.
Empirical values are taken from Ref.~\cite{DeVries:1987}.}
\label{fig:The-O16-Ca40-charge-densities}
\end{figure}

The charge density profile
is another important probe of nuclear structure. In Fig.~\ref{fig:The-O16-Ca40-charge-densities}
we show the charge densities of $^{16}$O and $^{40}$Ca calculated
with the pinhole algorithm. We have again taken into account the charge distribution
of the proton.
To compare with data from the electron scattering experiments we also
show results with the Coulomb interaction included via first
order perturbation theory. The Coulomb force suppresses the
central densities, drawing the results closer
to the empirical data. Our results are quite accurate for such a simple
nuclear interaction.

\begin{figure}[!ht]
\centering
\includegraphics[width=12cm]{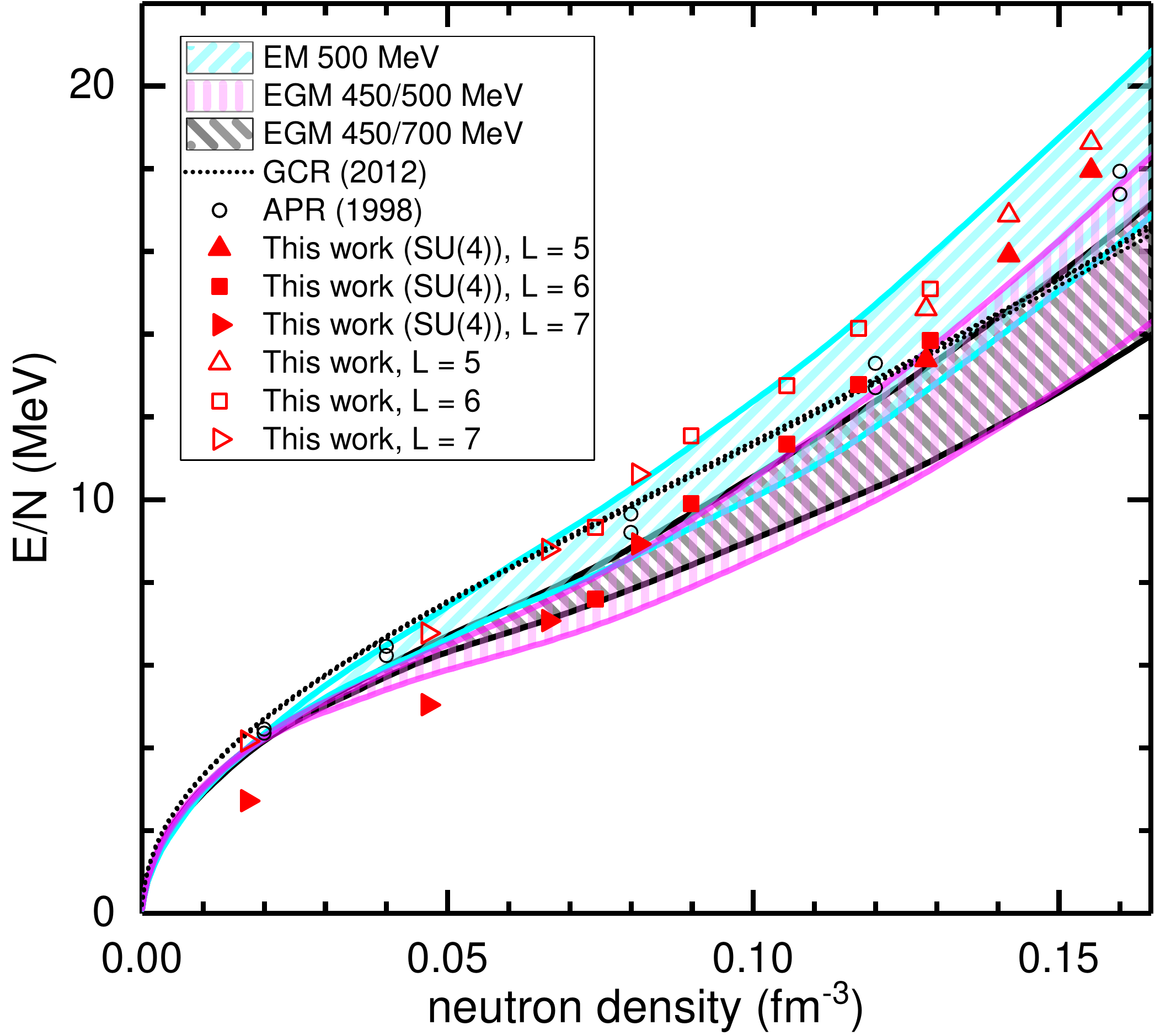}
\centering
\caption{{\bf Pure neutron matter.} The pure neutron matter
(NM) energy as a function of neutron density calculated using the NL50
interaction with box size $L$=5 (upright triangles), $L$=6 (squares), $L$=7 (rightward-pointing
triangles), respectively.  The filled red polygons show results for the 
leading-order SU(4)-symmetric interaction.  The open red 
polygons show an improved calculation with a short-range interaction 
to reproduce the physical neutron-neutron scattering length as well as a correction 
to improve invariance under Galilean boosts.  For comparison we also show results 
calculated with full N$^{3}$LO chiral
interactions
 (EM 500 MeV, EGM 450/500 MeV and EGM 450/700 MeV)~\cite{Tews:2012fj},
the results from variational (APR)~\cite{Akmal:1998cf} and Auxiliary Field Diffusion
MC calculations (GCR)~\cite{Gandolfi:2011xu}.}
\label{fig:The-calculated-pure-neutron-matter}
\end{figure}

We also examine the predictions for pure neutron matter (NM). In
Fig.~\ref{fig:The-calculated-pure-neutron-matter} we show the calculated
NM energy as a function of the neutron density and the comparison with other
calculations using next-to-next-to-next-to-leading-order (N$^{3}$LO) chiral
interactions.  In the lattice results we vary the number of neutrons from 14 to 66.  
The data for three different box sizes $L$=5 (upright triangles), $L$=6 (squares), $L$=7 (rightward-pointing
triangles) are marked as filled red polygons. 
We see that our results are in general agreement with the other calculations at densities
above 0.05~fm$^{-3}$, though calculations at higher orders are needed and are
planned in future work to estimate uncertainties.
At lower densities the discrepancy is larger as a result of our SU(4)-invariant
interaction having the incorrect neutron-neutron scattering length.  The open red 
polygons, again $L$=5 (upright triangles), $L$=6 (squares), $L$=7 (rightward-pointing
triangles), show an improved calculation with a short-range interaction 
to reproduce the physical neutron-neutron scattering length as well as a correction 
to improve invariance under Galilean boosts.  The restoration of Galilean invariance on 
the lattice is described in Ref.~\cite{Li:2019ldq}.  Overall, the results are quite 
good in view of the simplicity of the four-parameter interaction.

In this letter we have shown that the ground state
properties of light nuclei, medium-mass nuclei, and neutron matter can be
described using a minimal nuclear interaction with only four interaction
parameters.  While the first three parameters are
already standard in $\chi$EFT, the fourth
and last parameter is a new feature that controls the strength of the local
part of the nuclear interactions. These insights can help design $\chi$EFT interactions 
with better convergence at higher densities.  We encourage 
others to test simplified interactions in continuum nuclear structure calculations, interactions 
with SU(4) symmetry and a combination of local and nonlocal smearing.
The details of our interaction are given in the 
Supplemental Material.  In the continuum calculations, however, one can construct interactions 
with exact Galilean invariance, something that needs to be corrected order 
by order on the lattice \cite{Li:2019ldq}. 

We are now using SU(4)-symmetric short-range interactions with local and 
nonlocal smearing and also one-pion exchange as the starting point for improved 
calculations of light and medium-mass 
nuclei with chiral forces up to N$^3$LO.  While our ongoing N$^3$LO work is far from finished, we do know that the corrections at NLO are typically at the 10\% level in the binding energies.  We should clarify that what we called LO in lattice chiral effective field theory is actually an improved LO calculation where the S-wave effective range correction is included.  If the S-wave effective range correction were not included at LO, then the NLO correction would be at the 30\% level.  This 10\% correction at NLO might still seem too large since the agreement between the LO results in this work and the experimental binding energies are better than 10\%.  However, this better-than-expected agreement can be explained by the additional fine-tuning we gain by adjusting the balance between local and nonlocal interactions to achieve accurate liquid drop properties.

The main takeaway message of the work presented here is that while some fine tuning of 
the chiral forces seems 
necessary to improve convergence at higher densities, 
the number of independent fine tunings does not appear to be large.  While we have not solved the 
convergence problem, we characterized the scope of problem.  The key remaining 
question is how to accomplish these fine tunings without fitting 
to the many-body data that we wish to predict.  We plan to address this question in a 
forthcoming publication.

Aside from the Coulomb interaction, all of the other interactions in our
minimal model respect Wigner's SU(4) symmetry.  This is an example  of emergent
symmetry.  The SU(4)-invariant interaction resurges at higher densities not
because the underlying fundamental interaction is exactly invariant, but
because the SU(4)-invariant interaction is coherently enhanced in the many-body
environment.  This is not to minimize the important role of spin-dependent
effects such as spin-orbit couplings and tensor forces.  However, it does
seem to suggest that SU(4) invariance plays a key role in the bulk properties
of nuclear matter. 

The computational effort needed for the auxiliary-field lattice Monte Carlo
simulations scales with the number of nucleons, $A$, as somewhere between 
$A^1$ to $A^2$ for medium mass nuclei.  The actual exponent depends on the 
architecture of the computing platform.  The SU(4)-invariant interaction 
provides an enormous computational advantage by 
removing sign oscillations
from the lattice Monte Carlo simulations for any even-even nucleus.   Coulomb
interactions and all other corrections can be implemented using perturbation
theory or the recently-developed eigenvector continuation method if the corrections
are too large for perturbation theory~\cite{Frame:2017fah}. Given the mild
scaling with nucleon number and suppression of sign oscillations, the methods
presented here provide a new route to realistic lattice simulations of heavy
nuclei in the future with as many as one or two hundred
nucleons.  By realistic calculations we mean calculations where one can demonstrate order-by-order convergence in the chiral expansion going from LO to NLO, NLO to N$^2$LO, and N$^2$LO to N$^3$LO, while maintaining agreement with empirical data.

{\em
We thank A. Schwenk for providing the neutron matter results for comparison.
We acknowledge partial financial support from the Deutsche
Forschungsgemeinschaft (SFB/TRR 110, ``Symmetries and the Emergence
of Structure in QCD\textquotedblright ), the BMBF (Grant
No. 05P15PCFN1), the U.S. Department of Energy (DE-SC0018638 and DE-AC52-06NA25396),
and the Scientific and Technological Research Council of Turkey (TUBITAK
project no. 116F400). Further support was provided by the Chinese
Academy of Sciences (CAS) President\textquoteright s International
Fellowship Initiative (PIFI) (grant no. 2018DM0034) and by VolkswagenStiftung
(grant no. 93562). The computational resources were provided by the
Julich Supercomputing Centre at Forschungszentrum J\"ulich, Oak Ridge
Leadership Computing Facility, RWTH Aachen, and Michigan State University.
}

\clearpage

\renewcommand{\thefigure}{S\arabic{figure}}
\setcounter{figure}{0}
 
\section{Supplemental Material}

\subsection*{Auxiliary field formalism}

We simulate the interactions of nucleons on the lattice using projection
Monte Carlo with auxiliary fields; see Ref.~\cite{Lee:2008fa,Lahde:2019npb}
for an overview
of methods used in lattice EFT. We use an auxiliary-field formalism
where the interactions among nucleons are replaced by interactions
of nucleons with auxiliary fields at every lattice point in space
and time. In the auxiliary-field formalism each nucleon evolves as
if it is a single particle in a fluctuating background of auxiliary
fields. We use a single auxiliary field at LO in the EFT expansion
coupled to the total nucleon density. The interactions are reproduced
by integrating over the auxiliary field. In our lattice simulations,
the spatial lattice spacing is taken to be $a$ = (150 MeV)$^{-1}$=
1.32 fm, and the time lattice spacing is $a_{t}$ = (1000 MeV)$^{-1}$=
0.197 fm. For any fixed initial and final state, the amplitude for
a given configuration of auxiliary field is proportional to the determinant
of an $A\times A$ matrix $M_{ij}$. The entries of $M_{ij}$ are
the single nucleon amplitudes for a nucleon starting at state $j$
at $\tau$ = 0 and ending at state $i$ at $\tau$ = $\tau_{f}$.

We use a discrete auxiliary field that can simulate
the two-, three- and four-body forces simultaneously without sign oscillations.
To this end we write the interactions in the form,\begin{equation}
:\exp\left(-\frac{1}{2}Ca_{t}\rho^{2}-\frac{1}{6}C_{3}a_{t}\rho^{3}-\frac{1}{24}C_{4}a_{t}\rho^{4}\right):=\sum_{k=1}^{N}\omega_{k}:\exp\left(\sqrt{-Ca_{t}}\phi_{k}\rho\right):,\label{eq:Auxiliary_Field_Identity}
\end{equation}
where $C$ is the two-body coefficient, $C_{3}$ is three-body coefficient,
$C_{4}$ is the four-body coefficient, and the :: symbols indicate
the normal ordering of operators. We then solve for the real numbers
$\omega_{k}$ and $\phi_{k}$.  In this
work we only consider attractive two-body interactions with $C<0$.
In order to avoid the sign problem we further require $\omega_{k}>0$
for all $k$.

To determine the constants $\phi_{k}$ and $\omega_{k}$, we expand
Eq.~(\ref{eq:Auxiliary_Field_Identity}) up to $\mathcal{O}(\rho^{4})$
and compare both sides order by order. In the context of the nuclear
EFT, the three- and four-body interactions are usually much weaker
than the two-body interaction, and we use the following ansatz with $N=3$,
\begin{equation}
\omega_{1}=\frac{1}{\phi_{1}(\phi_{1}-\phi_{3})},\qquad\omega_{2}=1+\frac{1}{\phi_{1}\phi_{3}},\qquad\omega_{3}=\frac{1}{\phi_{3}(\phi_{3}-\phi_{1})}\label{eq:solutions}
\end{equation}
where $\phi_{2}=0$ and $\phi_{1}$ and $\phi_{3}$ are two
roots of the quadratic equation,
\begin{equation}
\phi^{2}+\frac{C_{3}}{\sqrt{-C^{3}a_{t}}}\phi-\frac{C_{3}^{2}}{C^{3}a_{t}}+\frac{C_{4}}{C^{2}a_{t}}-3=0.\label{eq:N3
value-1}
\end{equation}
Using Vieta's formulas relating polynomial coefficients to the sums and products
of roots, it is straightforward to verify that Eq.~(\ref{eq:solutions})
satisfies Eq.~(\ref{eq:Auxiliary_Field_Identity}) up to $\mathcal{O}(\rho^{4})$.
For a pure two-body interaction with $C_{3,4}=0$, the solution is simplified
to $\phi_{1}=-\phi_{3}=\sqrt{3}$, $\phi_{2}=0$, $\omega_{1}=\omega_{3}=1/6$,
$\omega_{2}=2/3$. The formalism Eq.~(\ref{eq:solutions}) is very
efficient for simulating the many-body forces. The
corresponding auxiliary field $s(n_{t},\bm{n})$ only assume three
different values $\phi_{1}$, $\phi_{2}$ and $\phi_{3}$ and can
be sampled with the shuttle algorithm described below.

\subsection*{Shuttle algorithm}

\begin{figure}[!ht]
\centering
\begin{centering}
\includegraphics[width=0.5\textwidth]{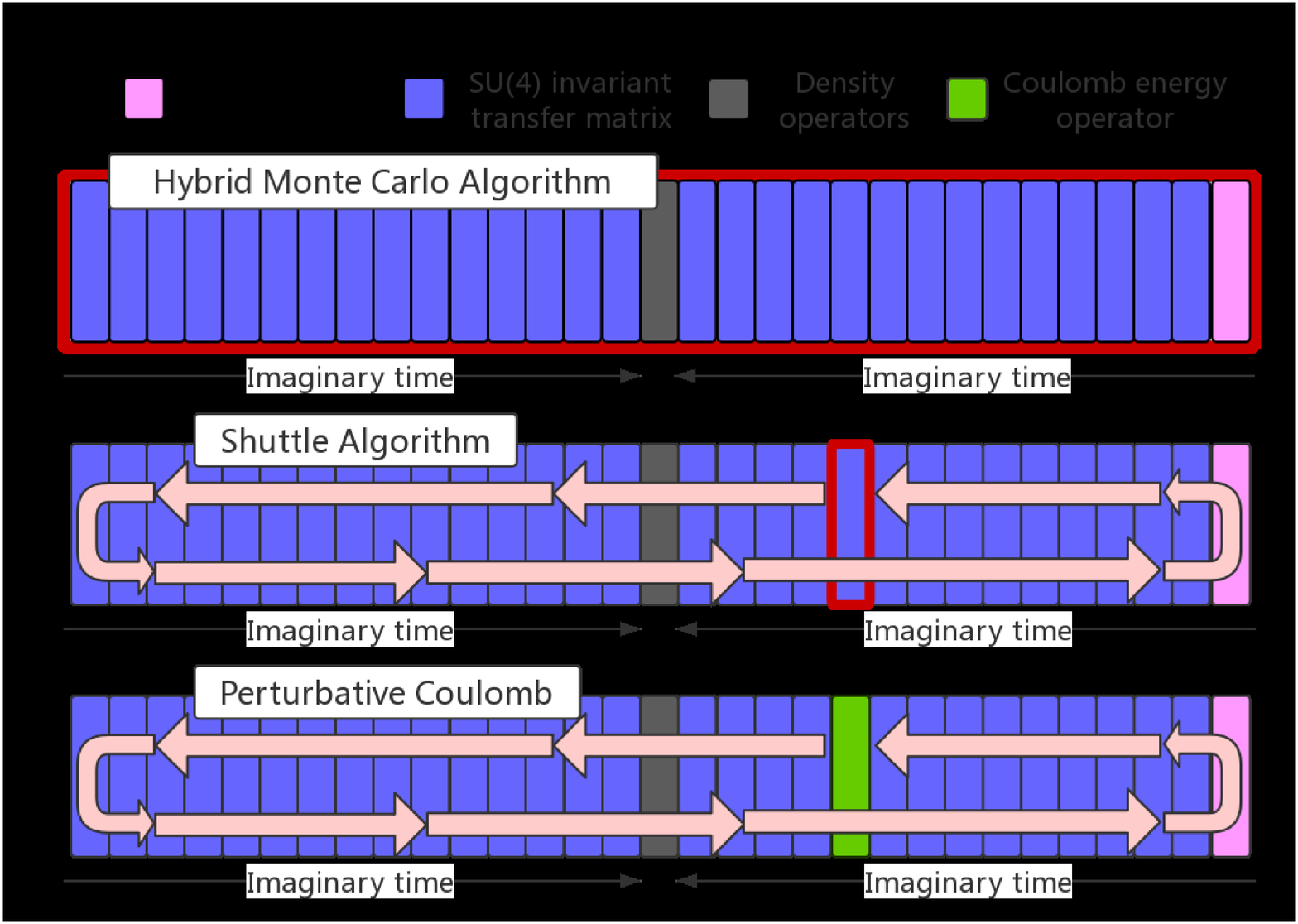}
\par\end{centering}
\caption{{\bf Update algorithms.} The schematic plot for the HMC algorithm
(upper panel), shuttle algorithm (middle panel) and the pinhole algorithm
with a perturbative Coulomb force. The red squares denote the time
slices to be updated in each run.}
\label{fig:The-schematic-plot}
\end{figure}

We update the auxiliary field $s(n_{t},\bm{n})$ using a shuttle algorithm
where only one time slice is updated at a time. In Fig.~\ref{fig:The-schematic-plot}
we show a schematic plot sketching the difference between the shuttle
algorithm and the Hybrid Monte Carlo (HMC) algorithm which performs an update
of all time slices. The shuttle
algorithm works as follows. 1) Choose one time slice $n_{t}$, record
the corresponding auxiliary field as $s_{{\rm old}}(n_{t},\bm{n}$).
2) Propose the new auxiliary fields $s_{{\rm new}}(n_{t},\bm{n}$) at each
lattice site
$\bm{n}$ according to the probability distribution $P\left[s_{{\rm new}}(n_{t},\bm{n})=\phi_{k}\right]=\omega_{k}$
for $k=1,2,3$. We note that $\omega_{1}+\omega_{2}+\omega_{3}=1$.  3) Calculate
the determinant of the $A \times A$ correlation matrix $M_{ij}$ using $s_{{\rm
old}}(n_{t},\bm{n})$
and $s_{{\rm new}}(n_{t},\bm{n})$, respectively. 4) Generate a random
number $r\in[0,1)$ and perform the following ``Metropolis test''. If
\[
\left|\frac{\det\left[M_{ij}\left(s_{{\rm new}}(n_{t},\bm{n})\right)\right]}{\det\left[M_{ij}\left(s_{{\rm
old}}(n_{t},\bm{n})\right)\right]}\right|>r,
\]
accept the new configuration $s_{{\rm new}}(n_{t},\bm{n})$ and update
the wave functions accordingly, otherwise keep $s_{{\rm old}}(n_{t},\bm{n})$.
5) Proceed to the next time slice, repeat steps 1)-4), and turn around at
the end of the time
series. As shown in Fig.\ref{fig:The-schematic-plot}, the program
runs back-and-forth like a shuttle bus and all the auxiliary fields are
updated after one cycle is finished.

The shuttle algorithm is well suited for small values of the temporal lattice
spacing $a_{t}$. In this
case the number of time slices is large and the impact of a single
update is small. In each update the new configuation is close to the
old one, resulting in a high acceptance rate. For example, in this work
the temporal lattice spacing is $a_{t}=0.001$ MeV$^{-1}$ and the
accept rate is around 50\% in most cases. We compared the
results with the HMC algorithm and found that the new algorithm is
more efficient. In most cases the number of independent configurations
per hour generated by the shuttle algorithm is three or four
times larger than that generated by the HMC algorithm.


\subsection*{Charge densities with Coulomb}

Auxiliary-field Monte Carlo simulations are efficient for computing
the quantum properties of systems with attractive pairing interactions.
By calculating the exact quantum amplitude for each configuration
of auxiliary fields, we obtain the full set of correlations induced
by the interactions. However, the exact quantum amplitude for each
auxiliary field configuration involves quantum states which are superpositions
of many different center-of-mass positions. Therefore information
about density correlations relative to the center of mass is lost.
Here we use the recently-developed pinhole algorithm to calculate
the charge density profiles in the center-of-mass frame. The details
of the algorithm can be found in Ref.~\cite{Elhatisari:2017eno}.

Let $\bm{n}_{k}=(\bm{r}_{k},s_{k},i_{k})$ denote the spatial coordinate,
spin and isospin of nucleon $k$. The one-body density at a point $\bm{r}$
in the intrinsic frame can
be written as
\begin{align}
\langle\rho(r)\rangle & =\frac{1}{A!}\sum_{\bm{n}_{1},\bm{n}_{2},\cdots,\bm{n}_{A}}\langle\Psi_{g.s.}|\rho_{A}|\Psi_{g.s.}\rangle\sum_{k=1}^{A}\delta(r-|\bm{r}_{k}-\bm{R}|)\label{eq:density_in_com}
\end{align}
where $\bm{R}=\frac{1}{A}\sum_{k}\bm{r}_{k}$ is the center of mass
of $A$ nucleons, $|\Psi_{g.s.}\rangle$ is the ground state, $\rho_{A}=:\rho(\bm{n}_{1})\rho(\bm{n}_{2})\cdots\rho(\bm{n}_{A}):$
is the $A$-body density operator. The summation over $\bm{n}$ means
a summation over all quantum numbers $\bm{n}_{1}$, $\bm{n}_{2}$,
$\cdots$, $\bm{n}_{A}$.

The ground state $|\Psi_{g.s.}\rangle$ can be rewritten using the
projection method,
\[
|\Psi_{g.s.}\rangle=\lim_{L_{t}\rightarrow\infty}M^{L_{t}}|\Psi\rangle,\qquad
M=:\exp(-a_{t}H):,
\]
where $H$ is the Hamiltonian, $M$ is the tranfer matrix. Then Eq.~(\ref{eq:density_in_com})
can be expressed using transfer matrices as
\begin{equation}
\langle\rho(r)\rangle=\frac{\sum_{\bm{n}}\langle\Psi|M^{L_{t}/2}\rho_{A}(\bm{n})M^{L_{t}/2}|\Psi\rangle\sum_{k=1}^{A}\delta(r-|\bm{r}_{k}-\bm{R}|)}{\sum_{\bm{n}}\langle\Psi|M^{L_{t}/2}\rho_{A}(\bm{n})M^{L_{t}/2}|\Psi\rangle}.\label{eq:density_in_com_at_LO}
\end{equation}
For the full Hamiltonian including Coulomb, Eq.~(\ref{eq:density_in_com_at_LO})
can not be computed directly using the Monte Carlo method because
the repulsive Coulomb interaction induces sign oscillations. To
solve this problem we employ perturbation theory. We split the
Hamiltonian into two parts,
\[
H=H^{(0)}+H^{(1)},
\]
where $H^{(0)}$ is the leading order Hamiltonian consisting of the
kinetic energy term and the SU(4)-invariant contact interactions,
and $H^{(1)}$ stands for the Coulomb interaction. In what follows we
only keep terms linear in $H^{(1)}$ or $M^{(1)}$ and omit all the
higher order terms. The transfer matrix can be
split similarly,
\[
M=:\exp(-a_{t}H):=:\exp(-a_{t}H^{(0)}):-a_{t}:\exp(-a_{t}H^{(0)})H^{(1)}:=M^{(0)}+M^{(1)}.
\]

Up to $\mathcal{O}(M^{(1)})$ the density Eq. (\ref{eq:density_in_com_at_LO})
can be written as
\begin{align*}
\langle\rho(r)\rangle & =\frac{\mathcal{M}_{\rho}^{(0)}+\mathcal{M}_{\rho}^{(1)}}{\mathcal{M}^{(0)}+\mathcal{M}^{(1)}}=\frac{\mathcal{M}_{\rho}^{(0)}}{\mathcal{M}^{(0)}}+\left(\frac{\mathcal{M}_{\rho}^{(1)}}{\mathcal{M}^{(0)}}-\frac{\mathcal{M}_{\rho}^{(0)}\mathcal{M}^{(1)}}{(\mathcal{M}^{(0)})^{2}}\right)+\mathcal{O}\left((\text{\ensuremath{\mathcal{M}}}^{(1)})^{2}\right),
\end{align*}
where the amplitudes $\mathcal{M}^{(0)}$ and $\mathcal{M}_{\rho}^{(0)}$
are defined as
\begin{align*}
\mathcal{M}^{(0)} & =\sum_{\bm{n}}\langle\Psi|(M^{(0)})^{L_{t}}|\Psi\rangle,\\
\mathcal{M}_{\rho}^{(0)} & =\sum_{\bm{n}}\langle\Psi|(M^{(0)})^{L_{t}/2}\rho_{A}(\bm{n})(M^{(0)})^{L_{t}/2}|\Psi\rangle\sum_{k=1}^{A}\delta(r-|\bm{r}_{k}-\bm{R}|).
\end{align*}
$\mathcal{M}^{(1)}$ and $\mathcal{M}_{\rho}^{(1)}$ are obtained
by substituting one of the transfer matrices $M^{(0)}$ in $\mathcal{M}^{(0)}$
and $\mathcal{M}_{\rho}^{(0)}$ by $M^{(1)}$ and adding up all $L_{t}$
possibilities,
\begin{align*}
\mathcal{M}^{(1)} & =\sum_{\bm{n}}\sum_{n_{t}=0}^{L_{t}/2-1}\langle\Psi|(M^{(0)})^{L_{t}-n_{t}-1}M^{(1)}(M^{(0)})^{n_{t}}|\Psi\rangle+{\rm
c.c.},\\
\mathcal{M}_{\rho}^{(1)} & =\sum_{\bm{n}}\sum_{n_{t}=0}^{L_{t}/2-1}\langle\Psi|(M^{(0)})^{L_{t}/2}\rho_{A}(\bm{n})(M^{(0)})^{L_{t}/2-n_{t}-1}M^{(1)}(M^{(0)})^{n_{t}}|\Psi\rangle\sum_{k=1}^{A}\delta(r-|\bm{r}_{k}-\bm{R}|)+{\rm
c.c.}.
\end{align*}
The four amplitudes $\mathcal{M}^{(0)}$, $\mathcal{M}^{(1)}$, $\mathcal{M}_{\rho}^{(0)}$
and $\mathcal{M}_{\rho}^{(1)}$ can be calculated using the auxiliary-field
formalism described above.

\subsection*{SU(4) breaking effects}

\begin{figure}[!ht]
\centering
\includegraphics[width=0.50\columnwidth]{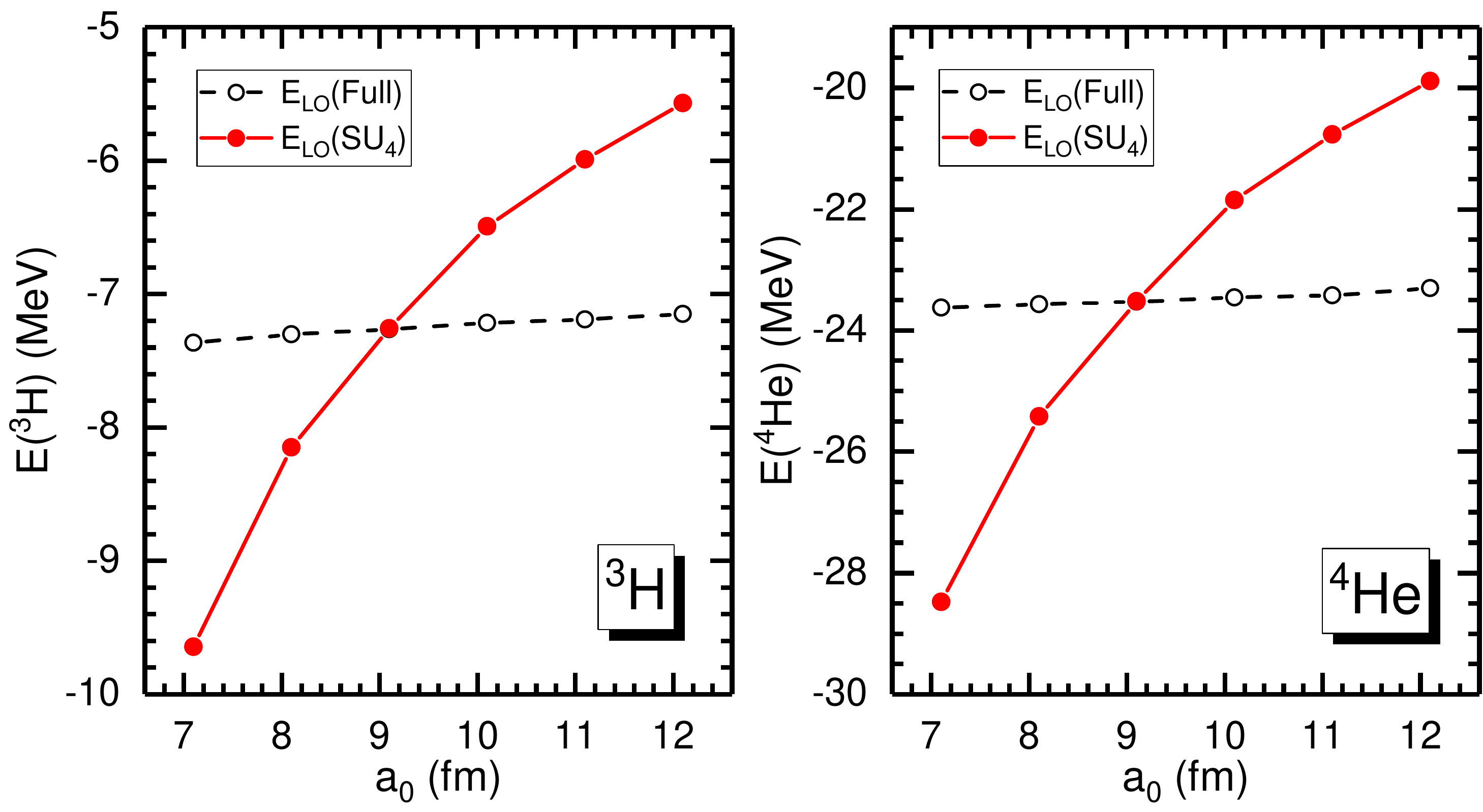}
\caption{{\bf Binding energies of $^{3}$H and $^{4}$He.} The solid symbols
denote
the binding energies of $^{3}$H and $^{4}$He calculated with leading order
interactions
fitted to different scattering length $a_0$ and fixed effective range $r_0=2.2$
fm.
The open symbols denote the results with the leading order SU(4) breaking
corrections fully included.
See the text for details.}
\label{fig:The-SU4-breaking-corrections}
\end{figure}

Our leading-order interactions fully respect Wigner's SU(4) symmetry, but
this symmetry is only approximate in nature. In order to optimize the strength
of our SU(4) interaction,
we calculate the energies of $^{3}$H and $^{4}$He using interactions corresponding
to different scattering lengths $a_0$
with fixed effective range $r_0=2.2$ fm. The results are shown in Fig.~\ref{fig:The-SU4-breaking-corrections}
as
full symbols. For each interaction we include the leading-order SU(4) breaking
effects for the two $S$-wave channels adjusted to reproduce the
experimental scattering lengths $a(^{1}$S${_0})$ and $a(^{3}$S${_1})$. The
corrected energies are calculated using
1st order perturbation theory and shown as open symbols.
The smaller values of $a_0$ correspond to stronger interactions and larger
binding energies. From Fig.~\ref{fig:The-SU4-breaking-corrections}
we can see immediately that the energy corrections for $^{3}$H and $^{4}$He
are both minimized at $a_0=9.1$ fm.
We take this value for all of our calculations.
This corresponds to a deuteron binding energy of $B(^2$H$) = 0.677$ MeV.
\subsection*{Volume and surface constants}

In this section we present the method for determining the parameter $s_{NL}$.
For each value of $s_{NL}$ we repeat entire process of fitting $a_0=9.1$
fm, $r_0=2.2$ fm,  and $B(^{3}$H$)=8.48$~MeV.
Each time this process results in different values for the local smearing
parameter $s_L$. We obtain
five such interactions with $s_{NL}=$0.40, 0.45, 0.50, 0.55 and 0.60
and denote them as NL40, NL45, NL50, NL55, and NL60, respectively. We note
that since
the effective range is kept constant, decreasing $s_{NL}$ corresponds to
increasing $s_L$ and thus the range of the local part of the interaction.
While we used alpha-alpha scattering to fix the local part of the interaction
in Ref.~\cite{Elhatisari:2016owd}, we are aware that such scattering calculations
are
difficult for other {\it ab initio} methods to reproduce. Therefore we adopt
a different approach that looks at the ground state energies of medium mass
nuclei.

For medium mass nuclei with $A\geq16$, the binding energies can
be well parameterized with the Bethe-Weizs{\"a}cker mass
formula,
\begin{equation}
B(A)=a_{V}A-a_{S}A^{\frac{2}{3}}+E_{{\rm Coulomb}}+\cdots,\label{eq:liquid_drop_mass_formula}
\end{equation}
where $a_{V}$ and $a_{S}$ are volume-energy and surface-energy constants,
respectively, $E_{{\rm Coulomb}}$ is the Coulomb energy, and the ellipsis
represents other terms such as the symmetry energy, pairing energy, shell
correction energy,
etc. To avoid fitting complexities not accurately captured in our minimal
nuclear interaction, we fit only $N=Z$ even-even nuclei,
for which the symmetry energy vanishes and the pairing energy varies smoothly.
The shell correction energy is known to be much smaller than
the macroscopic contribution in this mass region~\cite{Moller:1995} and
thus the first three terms appearing in Eq.~(\ref{eq:liquid_drop_mass_formula})
dominate.

\begin{figure}[!ht]
\centering
\includegraphics[width=0.50\columnwidth]{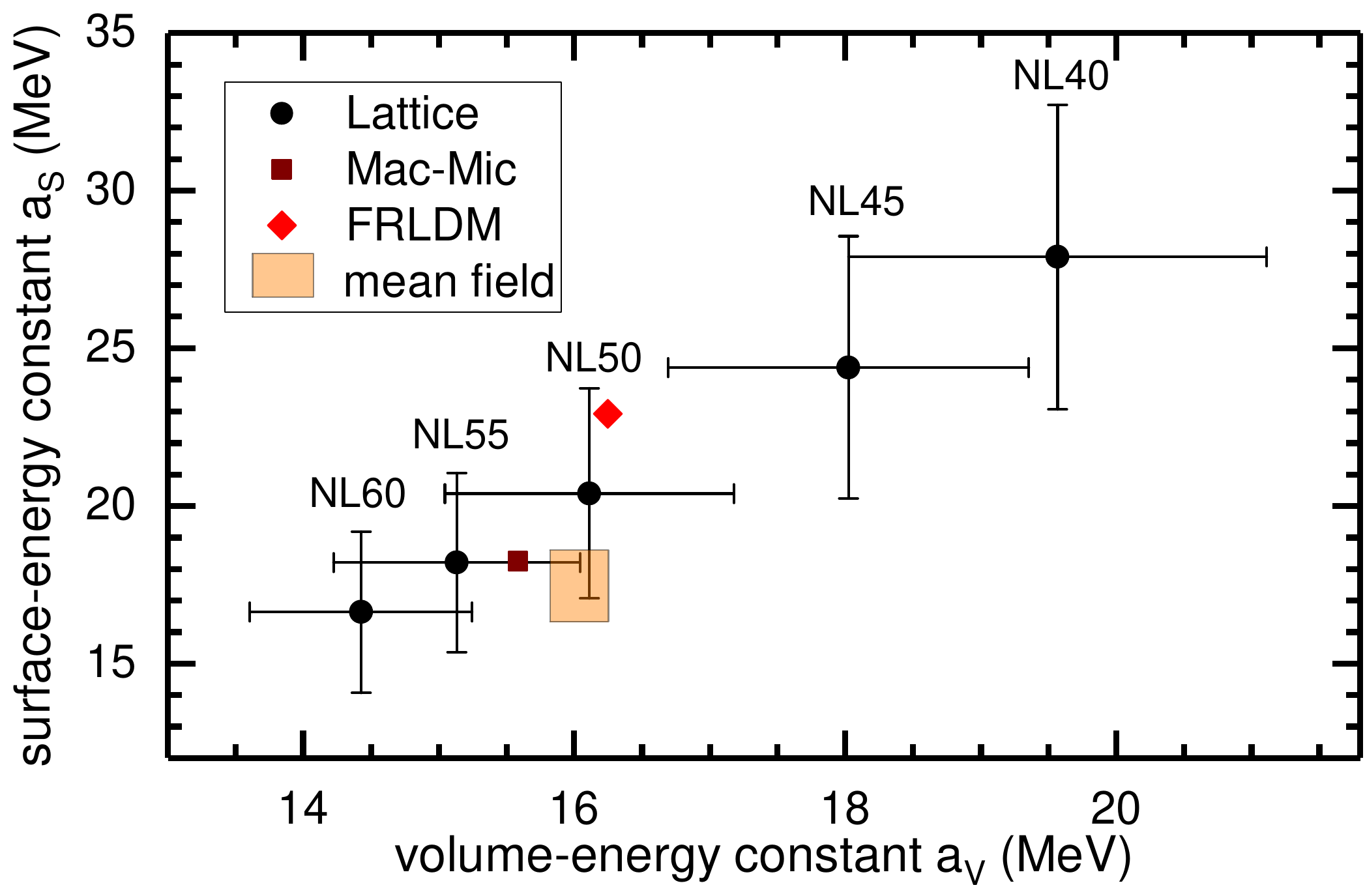}
\centering
\caption{{\bf Volume/Surface energy constants.} The correlation plot for
the calculated volume-energy constant $a_V$ and surface-energy constant $a_S$.
The square, diamond and square region denote the results fitted with
Macroscopic-Microscopic model~\cite{Wang:2014qqa}, Finite Range Liquid Drop
Model~\cite{Moller:1995}, Mean Field Models~\cite{Bender:2003jk}, respectively.}
\label{fig:The-correlation-plot-for-av-as}
\end{figure}

For each interaction we use the calculated binding energies with $16\leq
A\leq40$
to extract the liquid drop constants $a_{V}$ and $a_{S}$. We observe
prominent shell effects for these nuclei, and the binding energy
per nucleon fluctuates around the liquid drop values with maxima
at the magic numbers. In the fitting procedure the shell effects across
many nuclei averaged out, thus decreasing uncertainties for the
liquid drop constants.
The $a_{S}$-$a_{V}$ plot is shown in Fig.~\ref{fig:The-correlation-plot-for-av-as}.
We can see a linear correlation between these constants. The values of $a_{S}$
and $a_{V}$ both increase as the strength of the local part of the interaction
increases.  For comparison,
we also show other values of these constants in the literature where the
masses
are fitted throughout the entire chart of nuclides. We find that the interaction
NL50 gives a value of $a_{V}$ closest to the other estimations and corresponds
to about 16 MeV binding energy per nucleon at saturation.
The uncertainty
in $a_{S}$ is large but still matches the empirical values.

\subsection*{Data extrapolation and error analysis}

The Monte Carlo errors are calculated with a jackknife analysis. As
we employ a leading-order action free from sign oscillations, in most cases
the relative errors from the Monte Carlo simulation are smaller than
1\% and are not shown explicitly in the figures. The only exceptions
are the $^{40}$Ca charge density with Coulomb included shown in the
lower panel of Fig. 3, where the Monte Carlo errors become noticeable
at small radii.

In all the calculations in this letter we use a fixed number of temporal
steps
$L_{t}$=300. The exact ground state energies can be obtained by extrapolating
to the limit $L_{t}\rightarrow\infty$. To estimate the residual errors
from using a finite $L_{t}$, we perform multiple calculations with
$L_{t}$ varying from 100 to 300 for the nuclei listed in Table I.
The results are used to fit to the ansatz,
\[
E(L_{t})=E_{0}+c_{0}\exp\left(-\Delta EL_{t}a_{t}\right),
\]
where $E_{0}$, $c_{0}$ and $\Delta E$ are fit parameters. The differences
between the extrapolated energy $E_{0}$ and $E(L_{t}=300)$ are the
time extrapolation errors shown in Table I.  

\end{document}